\documentstyle[11pt]{article}
\begin{document}

\title{{\bf Surface Casimir densities on a spherical brane in Rindler-like
spacetimes}}
\author{ A. A. Saharian$^{1}$\thanks{%
E-mail: saharyan@server.physdep.r.am } and M. R. Setare$^{2}$\thanks{%
E-mail: rezakord@ipm.ir} \\
{\it $^1$ Department of Physics, Yerevan State University, Yerevan, Armenia }
\\
{\it $^2$ Institute for Theoretical Physics and Mathematics, Tehran, Iran}}
\maketitle

\begin{abstract}
The vacuum expectation value of the surface energy-momentum tensor
is evaluated for a scalar field obeying Robin boundary condition
on a spherical brane in $(D+1)$-dimensional spacetime $Ri\times
S^{D-1}$, where $Ri$ is a two-dimensional Rindler spacetime. The
generalized zeta function technique is used in combination with
the contour integral representation. The surface energies on
separate sides of the brane contain pole and finite contributions.
Analytic expressions for both these contributions are derived.
For an infinitely thin brane in odd spatial dimensions, the pole
parts cancel and the total surface energy, evaluated as the sum of
the energies on separate sides, is finite. For a minimally coupled
scalar field the surface energy-momentum tensor corresponds to the
source of the cosmological constant type.
\end{abstract}

\bigskip

PACS number(s): 03.70.+k, 11.10.Kk

\newpage

\section{Introduction}

\label{sec:Int}

Motivated by string/M theory, the AdS/CFT correspondence, and the
hierarchy problem of particle physics, braneworld models were
studied actively in recent years \cite{Hora96}. In this models,
our universe is realized as a boundary of a higher dimensional
spacetime. In particular, a well studied example is when the bulk
is an AdS space. The problem of studying quantum effects in
braneworld scenarios is of considerable phenomenological interest,
both in particle physics and in cosmology. The braneworld
corresponds to a manifold with dynamical boundaries and all fields
which propagate in the bulk will give Casimir-type contributions
to the vacuum energy, and as a result to the vacuum forces acting
on the branes. In dependence of the type of a field and boundary
conditions imposed, these forces can either stabilize or
destabilize the braneworld. In addition, the Casimir energy gives
a contribution to both the brane and bulk cosmological constants
and, hence, has to be taken into account in the self-consistent
formulation of the braneworld dynamics. Motivated by these, the
role of quantum effects on background of Randall--Sundrum geometry
has received a great deal of attention. The models with dS and AdS
branes, and higher dimensional brane models are considered as well
(see, for instance, references given in \cite{Saha05rs}).

In view of the recent developments in braneworld scenarios, it seems
interesting to generalize the study of quantum effects to other types of
bulk spacetimes. In particular, it is of interest to consider non-Poincar%
\'{e} invariant braneworlds, both to better understand the
mechanism of localized gravity and for possible cosmological
applications. Bulk geometries generated by higher-dimensional
black holes are of special interest. In these models , the tension
and the position of the brane are tuned in terms of black hole
mass and cosmological constant and brane gravity trapping occurs
in just the same way as in the Randall-Sundrum model. Braneworlds
in the background of the AdS black hole were studied in
\cite{AdSbhworld}. Like pure AdS space the AdS black hole may be
superstring vacuum. It is of interest to note that the phase
transitions which can be interpreted as confinement-deconfinement
transition in AdS/CFT setup may occur between pure AdS and AdS
black hole \cite{Witt98}. Though, in the generic black hole
background the investigation of brane-induced quantum effects is
technically complicated, the exact analytical results can be
obtained in the near horizon and large mass limit when the brane
is close to the black hole horizon. In this limit the black hole
geometry may be approximated by the Rindler-like manifold (for
some investigations of quantum effects on background of
Rindler-like spacetimes see \cite{Byts96} and references therein).
In the paper \cite{Saha05vol} we have investigated the Wightman
function, the vacuum expectation values of the field square and
the energy-momentum tensor for a scalar field with an arbitrary
curvature coupling parameter for the spherical brane on the bulk
$Ri\times S^{D-1}$, where $Ri$ is a two-dimensional Rindler
spacetime. This problem is also of separate interest as an example
with gravitational and boundary-induced polarizations of the
vacuum, where all calculations can be performed in a closed form.
Note that the corresponding quantities induced by a single and two
parallel flat branes in the bulk geometry $Ri\times R^{D-1}$ for
both scalar and electromagnetic fields are investigated in
\cite{Cand77}. For scalar fields with general curvature coupling,
in Ref. \cite{Rome02} it has been shown that in the discussion of
the relation between the mode sum energy, evaluated as the sum of
the zero-point energies for each normal mode of frequency, and the
volume integral of the renormalized energy density for the Robin
parallel plates geometry it is necessary to include in the energy
a surface term concentrated on the boundary (see also the
discussion in Ref. \cite{Full03}). An expression for the surface
energy-momentum tensor for a scalar field with a general curvature
coupling parameter in the general case of bulk and boundary
geometries is derived in Ref. \cite{Saha03}. The vacuum
expectation values of the surface energy-momentum tensor on the
branes in AdS bulk are investigated in \cite{Saha04surf}. In
particular, it has been shown that the surface densities induced
by quantum fluctuations of bulk fields can serve as a natural
mechanism for the generation of cosmological constant in
braneworld models of the Randall-Sundrum type with the value in
good agreement with recent cosmological observations. The purpose
of the present paper is to study the vacuum expectation value of
the surface energy-momentum tensor for a scalar field obeying
Robin boundary condition on a spherical brane on the bulk
$Ri\times S^{D-1}$.  The paper is organized as follows. In section
\ref{sec:surfemt} we consider the surface energy-momentum tensor
and the eigenfunctions for the problem. The vacuum expectation
value of the surface energy-momentum tensor in the R-region (the
definitions of the R- and L-regions see below) are investigated in
section \ref{sec:R-region}. The corresponding quantities for the
L-region are discussed in section \ref{sec:L-region}. Section
\ref{sec:Conc} summarizes the main results of the paper.

\section{Surface energy-momentum tensor}

\label{sec:surfemt}

Consider a real scalar field $\varphi (x)$ on background of $(D+1)$%
-dimensional spacetime $Ri\times S^{D-1}$, where $Ri$ is a two-dimensional
Rindler spacetime. The corresponding line element has the form%
\begin{equation}
ds^{2}=\xi ^{2}d\tau ^{2}-d\xi ^{2}-r_{H}^{2}d\Sigma _{D-1}^{2},
\label{ds22}
\end{equation}%
with the Rindler-like $(\tau ,\xi )$ part and $d\Sigma _{D-1}^{2}$ is the
line element for the space with positive constant curvature with the Ricci
scalar $R=(D-2)(D-1)/r_{H}^{2}$. Line element (\ref{ds22}) describes the
near horizon geometry of $(D+1)$-dimensional topological black hole with
coordinate $\xi $ determining the distance from the horizon. For example, in
the case of a $(D+1)$-dimensional Schwarzschild black hole one has $%
r-r_{H}=(D-2)\xi ^{2}/4r_{H}$, where $r$ is the Schwarzschild radial
coordinate and $r=r_{H}$ corresponds to the horizon. For the scalar field $%
\varphi (x)$ with curvature coupling parameter $\zeta $ the dynamics is
governed by the field equation
\begin{equation}
\left( \nabla _{l}\nabla ^{l}+m^{2}+\zeta R\right) \varphi =0,
\label{fieldeq}
\end{equation}%
where $\nabla _{l}$ is the covariant derivative operator
associated with the corresponding metric tensor $g_{ik}$. In the
cases of minimally and conformally coupled scalars one has $\zeta
=0$ and $\zeta =(D-1)/4D$, respectively. Our main interest in this
paper will be the surface Casimir energy and stresses induced on a
spherical brane located at $\xi =a$. We will assume that the field
satisfies the Robin boundary condition
\begin{equation}
(A_{s}+n^{l}\nabla _{l})\varphi (x)=0  \label{boundcond}
\end{equation}%
on the brane, where $A_{s}$ is a constant, $n^{l}$ is the unit inward normal
to the brane. This type of conditions is an extension of Dirichlet and
Neumann boundary conditions and appears in a variety of situations,
including the considerations of vacuum effects for a confined charged scalar
field in external fields \cite{Ambjorn2}, spinor and gauge field theories,
quantum gravity and supergravity \cite{Esp97}. Robin boundary conditions
naturally arise for scalar and fermion bulk fields in the Randall-Sundrum
model \cite{Gher00}. For boundary condition (\ref{boundcond}) the vacuum
expectation value of the bulk energy-momentum tensor induced by a spherical
brane is evaluated in Ref. \cite{Saha05vol}. In Ref. \cite{Saha03} it was
argued that the energy-momentum tensor for a scalar field on manifolds with
boundaries in addition to the bulk part contains a contribution located on
the boundary. The surface part of the energy-momentum tensor is given by the
formula \cite{Saha03}
\begin{equation}
T_{ik}^{{\rm (surf)}}=\delta (x;\partial M_{s})\tau _{ik}  \label{Ttausurf}
\end{equation}%
where the "one-sided" delta-function $\delta (x;\partial M_{s})$
locates this tensor on boundary $\partial M_{s}$ and
\begin{equation}
\tau _{ik}=\zeta \varphi ^{2}K_{ik}-(2\zeta -1/2)h_{ik}\varphi n^{l}\nabla
_{l}\varphi .  \label{tausurf}
\end{equation}%
Here $K_{ik}$ is the extrinsic curvature tensor for the boundary and $h_{ik}$
is the corresponding induced metric.

Let $\{\varphi _{\alpha }(x),\varphi _{\alpha }^{\ast }(x)\}$ be a complete
set of positive and negative frequency solutions to the field equation (\ref%
{fieldeq}), obeying boundary condition (\ref{boundcond}). Here $\alpha $
denotes a set of quantum numbers specifying the solution. By expanding the
field operator over the eigenfunctions $\varphi _{\alpha }(x)$, using the
standard commutation rules, for the vacuum expectation value of the surface
energy-momentum tensor one finds
\begin{equation}
\langle 0|T_{ik}^{{\rm (surf)}}|0\rangle =\delta (x;\partial M_{s})\langle
0|\tau _{ik}|0\rangle ,\quad \langle 0|\tau _{ik}|0\rangle =\sum_{\alpha
}\tau _{ik}\{\varphi _{\alpha }(x),\varphi _{\alpha }^{\ast }(x)\},
\label{modesumform}
\end{equation}%
where $|0\rangle $ is the amplitude for the vacuum state, and the bilinear
form $\tau _{ik}\{\varphi ,\psi \}$ on the right of the second formula is
determined by the classical energy-momentum tensor (\ref{tausurf}). To
evaluate the vacuum expectation value of the surface energy-momentum tensor
we need the eigenfunctions $\varphi _{\alpha }(x)$. In the consideration
below we will use the hyperspherical angular coordinates $(\vartheta ,\phi
)=(\theta _{1},\theta _{2},\ldots ,\theta _{n},\phi )$ on $S^{D-1}$\ with $%
n=D-2$, $0\leq \theta _{k}\leq \pi $, $k=1,\ldots ,n$, and $0\leq \phi \leq
2\pi $. In these coordinates the variables are separated and the
eigenfunctions can be written in the form%
\begin{equation}
\varphi _{\alpha }(x)=C_{\alpha }f(\xi )Y(m_{k};\vartheta ,\phi )e^{-i\omega
\tau },  \label{eigfunc1}
\end{equation}%
where $m_{k}=(m_{0}\equiv l,m_{1},\ldots m_{n})$, and $m_{1},m_{2},\ldots
m_{n}$ are integers such that%
\begin{equation}
0\leq m_{n-1}\leq \cdots \leq m_{1}\leq l,\quad -m_{n-1}\leq m_{n}\leq
m_{n-1},  \label{mk}
\end{equation}%
$Y(m_{k};\vartheta ,\phi )$ is the surface harmonic of degree $l$ \cite%
{Erdelyi}. The equation for $f(\xi )$ is obtained from field equation (\ref%
{fieldeq}). The corresponding linearly independent solutions are the Bessel
modified functions $I_{\pm i\omega }(\lambda _{l}\xi )$ and $K_{i\omega
}(\lambda _{l}\xi )$ with the imaginary order, where%
\begin{equation}
\quad \lambda _{l}=\frac{1}{r_{H}}\sqrt{l(l+n)+\zeta n(n+1)+m^{2}r_{H}^{2}}.
\label{lambdal}
\end{equation}%
The eigenfrequencies are determined from the boundary condition imposed on
the field at $\xi =a$. The brane divides the spacetime into two regions with
$\xi >a$ (R-region) and $0<\xi <a$ (L-region). The vacuum properties in
these regions are different and we consider them separately.

\section{Surface energy in the R-region}

\label{sec:R-region}

For the R-region the unit normal to the boundary and nonzero components of
the extrinsic curvature tensor have the form
\begin{equation}
n^{l}=\delta _{1}^{l},\quad K_{00}=a,  \label{normvec}
\end{equation}%
and $f(\xi )=K_{i\omega }(\lambda _{l}\xi )$. For a given $\lambda _{l}a$,
the corresponding eigenfrequencies $\omega =\omega _{j}=\omega _{j}(\lambda
_{l}a)$, $j=1,2,\ldots $, are determined from boundary condition (\ref%
{boundcond}) and are solutions to the equation
\begin{equation}
AK_{i\omega }(x)+xK_{i\omega }^{\prime }(x)=0,\quad x=\lambda _{l}a,\quad
A=A_{s}a,  \label{modeq}
\end{equation}%
where the prime denotes the differentiation with respect to the argument of
the function. For $A_{s}>0$ this equation has purely imaginary solutions
with respect to $\omega $. To avoid the vacuum instability, below we will
assume that $A_{s}\leq 0$. Under this condition all solutions to (\ref{modeq}%
) are real. The coefficient $C_{\alpha }$ in Eq. (\ref{eigfunc1}) is
determined by the normalization condition. Using the relation
\begin{equation}
\int \left\vert Y(m_{k};\vartheta ,\phi )\right\vert ^{2}d\Omega =N(m_{k})
\label{normsph}
\end{equation}%
for the spherical harmonics (the explicit form for $N(m_{k})$ will not be
necessary in the following consideration), one finds
\begin{equation}
C_{\alpha }^{2}=\frac{1}{r_{H}^{n+1}N(m_{k})}\frac{\bar{I}_{i\omega
_{j}}(\lambda _{l}a)}{\frac{\partial }{\partial \omega }\bar{K}_{i\omega
}(\lambda _{l}a)|_{\omega =\omega _{j}}},  \label{normcoef}
\end{equation}%
where for a given function $F(x)$ we use the notation
\begin{equation}
\bar{F}(x)=AF(x)+xF^{\prime }(x).  \label{barnot}
\end{equation}

Substituting the eigenfunctions into the mode-sum formula (\ref{modesumform}%
) and using the relations $K_{i\omega _{j}}(\lambda _{l}a)\bar{I}_{i\omega
_{j}}(\lambda _{l}a)=1$ and
\begin{equation}
\sum_{m_{k}}\frac{\left\vert Y(m_{k};\vartheta ,\phi )\right\vert ^{2}}{%
N(m_{k})}=\frac{D_{l}}{S_{D}},  \label{rel1}
\end{equation}%
the vacuum expectation value of the surface energy-momentum tensor can be
presented in the form
\begin{equation}
\langle 0|\tau _{l}^{k}|0\rangle =\frac{I_{{\rm R}}(A)}{2r_{H}^{D-1}aS_{D}}%
\left[ 2\zeta \delta _{l}^{0}\delta _{0}^{k}+(4\zeta -1)A\delta _{l}^{k}%
\right] ,\quad l,k=0,2,\ldots ,D,  \label{Tsurf}
\end{equation}%
and $\langle 0|\tau _{1}^{1}|0\rangle =0$, with $S_{D}=2\pi ^{D/2}/\Gamma
(D/2)$ being the total area of the surface of the unit sphere in $D$%
-dimensional space, and
\begin{equation}
I_{{\rm R}}(A)=\sum_{l=0}^{\infty }D_{l}\sum_{j=1}^{\infty }\frac{K_{i\omega
_{j}}(\lambda _{l}a)}{\frac{\partial }{\partial \omega }\bar{K}_{i\omega
}(\lambda _{l}a)|_{\omega =\omega _{j}}}.  \label{IA1}
\end{equation}%
Here and below the quantities for the R- and L-regions are denoted by the
indices R and L, respectively, and we use the notation%
\begin{equation}
D_{l}=(2l+D-2)\frac{\Gamma (l+D-2)}{\Gamma (D-1)l!}  \label{Dl}
\end{equation}%
for the degeneracy factor. The vacuum expectation value of the
surface energy-momentum tensor (\ref{Tsurf}) has a diagonal
structure:
\begin{equation}
\langle 0|\tau _{l}^{k}|0\rangle ={\rm diag}\left( \varepsilon ^{{\rm (R)}%
},0,-p^{{\rm (R)}},\ldots ,-p^{{\rm (R)}}\right) ,  \label{taudiag}
\end{equation}%
with the surface energy density $\varepsilon ^{{\rm (R)}}$, the stress%
\begin{equation}
p^{{\rm (R)}}=\frac{AI_{R}(A)}{2r_{H}^{D-1}a}(1-4\zeta ),  \label{surfstress}
\end{equation}
and with the equation of state%
\begin{equation}
\varepsilon ^{{\rm (R)}}=-\left[ 1+\frac{2\zeta }{A(4\zeta -1)}\right] p^{%
{\rm (R)}}.  \label{eqstate}
\end{equation}%
For a minimally coupled scalar field, the latter corresponds to a
cosmological constant induced on the brane. Note that the vacuum expectation
values of the field square on the brane is also expressed in terms of the
function $I_{R}(A)$:%
\begin{equation}
\langle 0|\varphi ^{2}|0\rangle _{\xi =a}=\frac{I_{R}(A)}{r_{H}^{D-1}S_{D}}.
\label{phi2on}
\end{equation}

The quantity (\ref{IA1}) and, hence, the surface energy-momentum tensor
diverges and needs some regularization. Many regularization techniques are
available nowadays and, depending on the specific physical problem under
consideration, one of them may be more suitable than the others. Here we
will use the method which is an analog of the generalized zeta function
approach. We define the function
\begin{equation}
F_{{\rm R}}(s)=\sum_{l=0}^{\infty }D_{l}\zeta _{{\rm R}}(s,\lambda _{l}a),
\label{IAs}
\end{equation}%
where
\begin{equation}
\zeta _{{\rm R}}(s,\lambda _{l}a)=\sum_{j=1}^{\infty }\frac{\omega
_{j}^{-s}K_{i\omega _{j}}(\lambda _{l}a)}{\frac{\partial }{\partial \omega }%
\bar{K}_{i\omega }(\lambda _{l}a)|_{\omega =\omega _{j}}}.  \label{zetsx}
\end{equation}%
Note that for Dirichlet boundary condition this function vanishes.
The computation of vacuum expectation value for the surface
energy-momentum tensor requires an analytical continuation of the
function $F_{{\rm R}}(s)$ to the value $s=0$,
\begin{equation}
I_{{\rm R}}(A)=F_{{\rm R}}(s)|_{s=0}.  \label{IFs0}
\end{equation}

The starting point of our consideration is the representation of the
function (\ref{zetsx}) in terms of contour integral
\begin{equation}
\zeta _{{\rm R}}(s,x)=\frac{1}{2\pi i}\int_{C}dz\,z^{-s}\frac{K_{iz}(x)}{%
\bar{K}_{iz}(x)},  \label{intzetsx1}
\end{equation}%
where $C$ is a closed counterclockwise contour in the complex $z$ plane
enclosing all zeros $\omega _{j}(x)$. The location of these zeros \ enables
one to deform the contour $C$ into a segment of the imaginary axis $(-iR,iR)$
and a semicircle of radius $R$ in the right half-plane. We will also assume
that the origin is avoided by the semicircle $C_{\rho }$ with small radius $%
\rho $. For sufficiently large $s$ the integral over the large semicircle in
(\ref{intzetsx1}) tends to zero in the limit $R\rightarrow \infty $, and the
expression on the right can be transformed to
\begin{equation}
\zeta _{{\rm R}}(s,x)=\frac{1}{2\pi i}\int_{C_{\rho }}dz\,z^{-s}\frac{%
K_{iz}(x)}{\bar{K}_{iz}(x)}-\frac{1}{\pi }\cos \frac{\pi s}{2}\int_{\rho
}^{\infty }dz\,z^{-s}\frac{K_{z}(x)}{\bar{K}_{z}(x)}.  \label{intzetsx2}
\end{equation}%
Below we will consider the limit $\rho \rightarrow 0$. In this limit the
first integral vanishes in the case $s=0$, and in the following we will
concentrate on the contribution of the second integral. For the analytic
continuation of this integral we employ the uniform asymptotic expansion of
the MacDonald function and its derivative for large values of the order \cite%
{Abramowitz}. We will rewrite this expansion in the form
\begin{equation}
K_{z}(x)\sim \sqrt{\frac{\pi }{2}}\frac{e^{-z\eta (x/z)}}{(x^{2}+z^{2})^{1/4}%
}\sum_{q=0}^{\infty }\frac{(-1)^{q}\widetilde{u}_{q}(t)}{(x^{2}+z^{2})^{q/2}}%
,  \label{Kzxasymp}
\end{equation}%
where
\begin{equation}
t=\frac{z}{\sqrt{x^{2}+z^{2}}},\quad \eta (x)=\sqrt{1+x^{2}}+\ln \frac{x}{1+%
\sqrt{1+x^{2}}},\quad \tilde{u}_{q}(t)=\frac{u_{q}(t)}{t^{q}},
\label{tetaul}
\end{equation}%
and the expressions for the functions $u_{q}(t)$ are given in \cite%
{Abramowitz}. From these expressions it follows that the coefficients $%
\tilde{u}_{q}(t)$ have the structure
\begin{equation}
\tilde{u}_{q}(t)=\sum_{m=0}^{q}u_{qm}t^{2m},  \label{ult}
\end{equation}%
with numerical coefficients $u_{qm}$. From Eq. (\ref{Kzxasymp}) and the
corresponding expansion for the derivative of the MacDonald function we
obtain the asymptotic expansion
\begin{equation}
\bar{K}_{z}(x)\sim -\sqrt{\frac{\pi }{2}}(x^{2}+z^{2})^{1/4}e^{-z\eta
(x/z)}\sum_{q=0}^{\infty }\frac{(-1)^{q}\tilde{v}_{q}(t)}{(x^{2}+z^{2})^{q/2}%
}\,,  \label{Kzxbaras}
\end{equation}%
where
\begin{equation}
\tilde{v}_{q}(t)=\frac{v_{q}(t)}{t^{q}}+A\tilde{u}_{q-1}\,,  \label{vltbar}
\end{equation}%
and the expressions for $v_{q}(t)=t^{q}\sum_{m=0}^{q}v_{qm}t^{2m}$ are
presented in \cite{Abramowitz}. Note that the functions (\ref{vltbar}) have
the structure
\begin{equation}
\tilde{v}_{q}(t)=\sum_{m=0}^{q}\tilde{v}_{qm}t^{2m},\quad \tilde{v}%
_{qm}=v_{qm}+Au_{q-1,m}\,.  \label{vltbar1}
\end{equation}%
From Eqs. (\ref{Kzxasymp}) and (\ref{Kzxbaras}) we can find the asymptotic
expansion for the ratio in the second integral on the right of formula (\ref%
{intzetsx2}):
\begin{equation}
\frac{K_{z}(x)}{\bar{K}_{z}(x)}\sim -\frac{1}{(x^{2}+z^{2})^{1/2}}%
\sum_{q=0}^{\infty }\frac{(-1)^{q}U_{q}(t)}{(x^{2}+z^{2})^{q/2}}\,,
\label{Kzratio}
\end{equation}%
where the coefficients $U_{q}(t)$ are defined by the relation
\begin{equation}
\sum_{q=0}^{\infty }(-1)^{q}\frac{\tilde{u}_{q}(t)}{r^{q}}\left[
\sum_{q=0}^{\infty }(-1)^{q}\frac{\tilde{v}_{q}(t)}{r^{q}}\right]
^{-1}=\sum_{q=0}^{\infty }\frac{(-1)^{q}U_{q}(t)}{r^{q}}\,,  \label{defUl}
\end{equation}%
and similar to (\ref{ult}), (\ref{vltbar1}), are polynomials in $t$:%
\begin{equation}
U_{q}(t)=\sum_{j=0}^{q}U_{qj}t^{2j}.  \label{Ulmdef}
\end{equation}%
The first three coefficients are given by expressions
\begin{eqnarray}
U_{0}(t) &=&1,\;U_{1}(t)=\frac{1}{2}-A-\frac{t^{2}}{2},  \nonumber
\label{Ufunc} \\
U_{2}(t) &=&\frac{3}{8}-A+A^{2}-\left( A-\frac{7}{32}\right) t^{2}+\frac{49}{%
576}\,t^{4}.  \nonumber
\end{eqnarray}

Now let us consider the function
\begin{equation}
F_{{\rm R}}(s)=-\frac{1}{\pi }\cos \frac{\pi s}{2}\sum_{l=0}^{\infty
}D_{l}\int_{\rho }^{\infty }dz\,z^{-s}\frac{K_{z}(\lambda _{l}a)}{\bar{K}%
_{z}(\lambda _{l}a)}.  \label{Fs}
\end{equation}%
We subtract and add to the integrand in this equation the first $N$ terms of
the corresponding asymptotic expansion. This allows us to split (\ref{Fs})
into the following pieces
\begin{equation}
F_{{\rm R}}(s)=F_{{\rm R}}^{(as)}(s)+F_{{\rm R}}^{(1)}(s)\,,  \label{FasF1}
\end{equation}%
where
\begin{eqnarray}
F_{{\rm R}}^{(as)}(s) &=&\frac{1}{\pi }\cos \frac{\pi s}{2}%
\sum_{l=0}^{\infty }D_{l}\int_{\rho }^{\infty }dz\,z^{-s}\sum_{q=0}^{N}\frac{%
(-1)^{q}U_{q}(t)}{(z^{2}+\lambda _{l}^{2}a^{2})^{(q+1)/2}},  \label{Fas} \\
F_{{\rm R}}^{(1)}(s) &=&-\frac{1}{\pi }\cos \frac{\pi s}{2}%
\sum_{l=0}^{\infty }D_{l}\int_{\rho }^{\infty }dz\,z^{-s}\left[ \frac{%
K_{z}(\lambda _{l}a)}{\bar{K}_{z}(\lambda _{l}a)}+\sum_{q=0}^{N}\frac{%
(-1)^{q}U_{q}(t)}{(z^{2}+\lambda _{l}^{2}a^{2})^{(q+1)/2}}\right] ,
\label{F1}
\end{eqnarray}%
and
\begin{equation}
t=z/\sqrt{z^{2}+\lambda _{l}^{2}a^{2}}.  \label{rtet}
\end{equation}%
For $N\geq D-1$ the expression for $F_{{\rm R}}^{(1)}(s)$ is finite at $s=0$
and, hence, for our aim it is sufficient to subtract $N=D-1$ asymptotic
terms. At $s=0$ the function $F_{{\rm R}}^{(1)}(s)$ is finite for $\rho =0$
and we can directly put this value. The integral over $z$ in the expression
for $F_{{\rm R}}^{(as)}(s)$ is finite in the limit $\rho \rightarrow 0$ for $%
0<{\rm Re\,}s<1$. For these values we can put $\rho =0$ in Eq. (\ref{Fas}).
By making use formulae (\ref{Ulmdef}), (\ref{rtet}), after the integration
over $z$, the asymptotic part is presented in the form%
\begin{equation}
F_{{\rm R}}^{(as)}(s)=\frac{1}{2\pi }\cos \frac{\pi s}{2}%
\sum_{q=0}^{N}(-1)^{q}\left( \frac{r_{H}}{a}\right)
^{q-s}\sum_{j=0}^{q}U_{qj}B\left( j+\frac{1-s}{2},\frac{q+s}{2}\right) \zeta
_{S^{D-1}}\left( \frac{q+s}{2}\right) ,  \label{Faszintheto}
\end{equation}%
with the beta function $B(x,y)$. In formula (\ref{Faszintheto})
\begin{equation}
\zeta _{S^{D-1}}(z)==\sum_{l=0}^{\infty }D_{l}\left[ (l+D/2-1)^{2}+b_{D}%
\right] ^{-z},  \label{zetaS}
\end{equation}%
is the zeta function for a scalar field on the spacetime $R\times S^{D-1}$
and
\begin{equation}
b_{D}=\zeta (D-2)(D-1)-(D-2)^{2}/4+m^{2}r_{H}^{2}.  \label{bn}
\end{equation}%
This function is well investigated in literature (see, for example, \cite%
{Camp90}) and can be presented as a series of incomplete zeta
functions. Here we recall that the function $\zeta _{S^{D-1}}(z)$
is a meromorphic function with simple poles at $z=(D-1)/2-j$,
where $j=0,1,2,\ldots $ for $D$ even and $0\leq j\leq (D-3)/2$ for
$D$ odd. For $D$ even one has $\zeta _{S^{D-1}}(-j)=0$,
$j=1,2,\ldots $. In (\ref{Faszintheto}), the pole term in the
$q=0$ summand comes from the pole of the beta function, whereas in
the terms with $q\neq 0$ the pole terms come from the poles of the
function $\zeta _{S^{D-1}}(z)$. Laurent-expanding near $s=0$ we
find
\begin{equation}
F_{{\rm R}}(s)=\frac{F_{{\rm R},-1}^{(as)}}{s}+F_{{\rm R},0}^{(as)}+F_{{\rm R%
}}^{(1)}(0)+{\cal {O}}(s).  \label{Fas4}
\end{equation}%
Using this result, for the surface energy density induced on the brane one
obtains%
\begin{equation}
p^{{\rm (R)}}=p_{p}^{{\rm (R)}}+p_{f}^{{\rm (R)}},  \label{ppf}
\end{equation}%
where for the pole and finite contributions one has%
\begin{eqnarray}
\varepsilon _{p}^{{\rm (R)}} &=&\frac{A(4\zeta -1)+2\zeta }{%
2sr_{H}^{D-1}aS_{D}}F_{{\rm R},-1}^{(as)},  \label{ppf1} \\
\varepsilon _{f}^{{\rm (R)}} &=&\frac{A(4\zeta -1)+2\zeta }{%
2r_{H}^{D-1}aS_{D}}\left[ F_{{\rm R},0}^{(as)}+F_{{\rm R}}^{(1)}(0)\right] .
\end{eqnarray}%
The corresponding formulae for the pole and finite parts of the surface
stress are obtained by using the equation of state (\ref{eqstate}). The
surface energy can be found integrating the energy density,%
\begin{equation}
E^{{\rm (R,surf)}}=\int d^{D}x\sqrt{|g|}\langle 0|T_{0}^{{\rm (surf)}%
0}|0\rangle =ar_{H}^{D-1}S_{D}\varepsilon ^{{\rm (R)}}.  \label{ERsurf}
\end{equation}%
The pole and finite parts of the vacuum expectation value of the field
square on the brane are obtained by the formulae (\ref{phi2on}), (\ref{IFs0}%
), (\ref{Fas4}).

\section{Surface densities in the L-region}

\label{sec:L-region}

In this section we consider the region between the horizon and the
brane, $0<\xi <a$ (L-region), for which one has $n^{l}=-\delta
_{1}^{l}$ and $K_{00}=-a$. As in the previous section we will
assume that the field obeys boundary condition (\ref{boundcond})
on the surface $\xi =a$. To deal with discrete spectrum, we can
introduce the second brane located at $\xi =b<a$, on whose surface
we impose boundary conditions as well. After the construction of
the corresponding zeta function we take the limit $b\rightarrow
0$. As a result, we can see that the
surface energy-momentum tensor in the L-region has the structure given by (%
\ref{taudiag}) and with the equation of state (\ref{eqstate}). For the
surface energy density one obtains the expression
\begin{equation}
\varepsilon ^{{\rm (L)}}=\frac{A(4\zeta -1)+2\zeta }{2r_{H}^{D-1}aS_{D}}I_{%
{\rm L}}(A),\quad A=-aA_{s},  \label{surfstressL}
\end{equation}%
where now $I_{{\rm L}}(A)=F_{{\rm L}}(s)|_{s=0}$ with
\begin{equation}
F_{{\rm L}}(s)=-\frac{1}{\pi }\cos \frac{\pi s}{2}\sum_{l=0}^{\infty
}D_{l}\int_{\rho }^{\infty }dz\,z^{-s}\frac{I_{z}(\lambda _{l}a)}{\bar{I}%
_{z}(\lambda _{l}a)}.  \label{IRL}
\end{equation}%
For a given $A$ this expression differs from the corresponding expression
for the R-region by the replacement $K_{z}(x)\rightarrow I_{z}(x)$. Note
that the similar relation takes place for the bulk energy-momentum tensor as
well. As in the previous section, to avoid the vacuum instability, here we
assume that $A_{s}\leq 0$. Under this condition, for a given $\lambda _{l}a$
the function $\bar{I}_{z}(\lambda _{l}a)$ has no real positive zeros with
respect to $z$. The uniform asymptotic expansion for the integrand in (\ref%
{IRL}) is obtained from the corresponding formula with the functions $%
K_{z}(\lambda _{l}a)$ (see formula (\ref{Kzratio})) by the replacement%
\begin{equation}
(-1)^{q}U_{q}(t)\rightarrow -U_{q}(t).  \label{replace1}
\end{equation}%
The vacuum stress is a sum of pole and finite parts%
\begin{equation}
\varepsilon ^{{\rm (L)}}=\varepsilon _{p}^{{\rm (L)}}+\varepsilon _{f}^{{\rm %
(L)}},  \label{ppfL}
\end{equation}%
with%
\begin{eqnarray}
\varepsilon _{p}^{{\rm (L)}} &=&\frac{A(4\zeta -1)+2\zeta }{%
2sr_{H}^{D-1}aS_{D}}F_{{\rm L},-1}^{(as)},  \label{ppf1L} \\
\varepsilon _{f}^{{\rm (L)}} &=&\frac{A(4\zeta -1)+2\zeta }{%
2r_{H}^{D-1}aS_{D}}\left[ F_{{\rm L},0}^{(as)}+F_{{\rm L}}^{(1)}(0)\right] .
\nonumber
\end{eqnarray}%
The formulae for $F_{{\rm L},-1}^{(as)}$, $F_{{\rm L},0}^{(as)}$, $F_{{\rm L}%
}^{(1)}(0)$ are obtained from the corresponding expressions for the R-region
by the replacements $K_{z}(x)\rightarrow I_{z}(x)$ and (\ref{replace1}). In
particular,%
\begin{equation}
F_{{\rm L}}^{(as)}(s)=-\frac{1}{2\pi }\cos \frac{\pi s}{2}%
\sum_{q=0}^{N}\left( \frac{r_{H}}{a}\right)
^{q-s}\sum_{j=0}^{q}U_{qj}B\left(
j+\frac{1-s}{2},\frac{q+s}{2}\right) \zeta _{S^{D-1}}\left(
\frac{q+s}{2}\right) .
\end{equation}%
The surface energy density is related to the stress by formula (\ref{eqstate}%
) with the replacement R$\rightarrow $L and for the total surface energy one
has%
\begin{equation}
E^{{\rm (L,surf)}}=ar_{H}^{D-1}S_{D}\varepsilon ^{{\rm (L)}}.  \label{ELsurf}
\end{equation}%
The vacuum expectation value of the field square on the brane for
the L-region is also expressed in terms of the function
$I_{{\mathbf{L}}}(A)$. For an infinitely thin brane taking the R-
and L-regions together, the pole parts of the surface energy
densities cancel for odd values of the spatial dimension $D$.
In this case the total surface energy $E^{{\rm (surf)}}=E^{%
{\rm (R,surf)}}+E^{{\rm (L,surf)}}$ is finite and can be directly evaluated
by the formula%
\begin{eqnarray}
E^{{\rm (surf)}} &=&\frac{A(1-4\zeta )+2\zeta }{2\pi }\left\{
\sum_{k=0}^{N_{1}}\left( \frac{r_{H}}{a}\right) ^{2k+1}\zeta
_{S^{D-1}}\left( k+\frac{1}{2}\right) \sum_{j=0}^{2k+1}U_{2k+1,j}B\left( j+%
\frac{1}{2},k+\frac{1}{2}\right) \right.   \nonumber \\
&&+\left. \sum_{l=0}^{\infty }D_{l}\int_{0}^{\infty }dz\,\left[ \frac{%
I_{z}(\lambda _{l}a)}{\bar{I}_{z}(\lambda _{l}a)}+\frac{K_{z}(\lambda _{l}a)%
}{\bar{K}_{z}(\lambda _{l}a)}-\sum_{k=0}^{N_{1}}\frac{2U_{2k+1}(t)}{%
(z^{2}+\lambda _{l}^{2}a^{2})^{k+1}}\right] \right\} ,  \label{epstot}
\end{eqnarray}%
where $N_{1}=[(N-1)/2]$, $N\geq D-1$, and $t$ is defined by relation (\ref%
{rtet}). Note that the cancellation of the pole terms coming from
oppositely oriented faces of infinitely thin smooth boundaries
takes place in vary many situations encountered in the literature.
It is a simple consequence of the fact that the second fundamental
forms are equal and opposite on two faces of each boundary and,
consequently, the values of the corresponding coefficient in the
heat kernel expansion summed over two faces of each boundary
vanishes.

We have investigated the surface densities for both R- and
L-regions. In the corresponding braneworld scenario the geometry
is made up by two slices of the region $0<\xi <a$ glued together
at the brane with a orbifold-type symmetry condition analogous to
that in the Randall-Sundrum model (see, for instance,
\cite{Saha05vol}). For an untwisted scalar field the coefficient
$A_{s}$ in the boundary condition is related to the brane mass
parameter $c$ of the field and the extrinsic curvature of the
brane by the relation $A_{s}=(c-\zeta /a)/2$. For a twisted scalar
Dirichlet boundary condition is obtained. It should be noted that
in the orbifolded version due to $Z_2$ symmetry the extrinsic
curvature tensor is the same on both sides of the fixed point and
the cancellation of the pole terms for odd values $D$ does not
take place. A natural way to deal with surface divergences is to
consider more realistic brane models with finite thickness. As it
has been discussed in \cite{Mina06} for de Sitter brane model, the
finite thickness of the brane regularizes the ultraviolet behavior
and acts as a natural cutoff.

\section{Conclusion}

\label{sec:Conc}

In this paper we have investigated the surface Casimir densities induced on
a spherical brane in the Rindler-like spacetime $Ri\times S^{D-1}$ by
quantum fluctuations of a scalar field with an arbitrary curvature coupling
parameter. The corresponding volume vacuum expectation values of the
energy--momentum tensor were investigated in \cite{Saha05vol}. We consider a
scalar field with Robin boundary conditions and as a regularization method
the zeta function technique is employed. The spherical brane divides the
background space into two regions, referred as R- and L-regions. We have
constructed an integral representations for the corresponding zeta functions
in both these regions, which are well suited for the analytic continuation.
Subtracting and adding to the integrands the leading terms of the
corresponding uniform asymptotic expansions, we present the corresponding
functions as a sum of two parts. The first one is convergent at the physical
point and can be evaluated numerically. In the second, asymptotic part the
pole contributions are given explicitly in terms of the zeta function for a
scalar field on the spacetime $R\times S^{D-1}$. The latter is
well-investigated in literature. As a consequence, the vacuum expectation
values of the surface energy-momentum tensor for separate R- and L-regions
contain pole and finite contributions. The remained pole term is a
characteristic feature for the zeta function regularization method and has
been found for many other cases of boundary geometries. For a minimally
coupled scalar field, the surface energy-momentum tensor induced by quantum
vacuum effects corresponds to a source of a cosmological constant type
located on the brane. In odd spatial dimensions in the case of an infinitely
thin brane, taking the R- and L-regions together, the pole parts of the
surface vacuum energies cancel. As a result the total surface energy is
finite and is determined by formula (\ref{epstot}) with the function $%
U_{q}(t)$ is defined by relation (\ref{defUl}). The results
obtained here can be applied to the braneworld in the AdS black
hole bulk in the limit when the brane is close to the black hole
horizon. In this paper we have considered the surface
energy-momentum tensor on a codimension one smooth brane. For
non-smooth boundaries an additional part in the energy-momentum
tensor arises located on corners. The corresponding corner terms
can be important in codimension two braneworld scenarios (see, for
instance, \cite{Agha03} and references therein).

\section*{Acknowledgement}

The work of A.A. Saharian has been supported by ANSEF Grant No.
05-PS-hepth-89-70 and in part by the Armenian Ministry of Education and
Science, Grant No.~0124.

\end{document}